\documentstyle[12pt,epsfig]{article} 
\newcommand{\NP}[1]{{\it Nucl.\ Phys.}\ {\bf #1}}

\newcommand{\PL}[1]{{\it Phys.\ Lett.}\ {\bf #1}} 
\newcommand{\PR}[1]{{\it Phys.\ Rev.}\ {\bf #1}} 
\newcommand{\PRL}[1]{{\it Phys.\ Rev.\ Lett.}\ {\bf #1}} 
\newcommand{\IJMP}[1]{{\it Int.\ J.\ Mod.\ Phys.}\ {\bf #1}} 
\newcommand{\MPL}[1]{{\it Mod.\ Phys.\ Lett.}\ {\bf #1}}

\newcommand{\bfp}{\mbox{\boldmath $p$}} 
\newcommand{\bfP}{\mbox{\boldmath $P$}} 
\newcommand{\bfk}{\mbox{\boldmath $k$}}

\newcommand{\simorder}{\raisebox{-4pt}{$\, \stackrel{\textstyle >}{\sim} \,$}}

\newcommand{\Lup}{\Lambda^\uparrow} 
\newcommand{\Ldown}{\Lambda^\downarrow}

\newcommand{\hup}{h^\uparrow} 
\newcommand{\hdown}{h^\downarrow} 
\newcommand{\beq}{\begin{equation}} 
\newcommand{\eeq}{\end{equation}} 
\newcommand{\pup}{p^\uparrow} 
\newcommand{\aup}{a^\uparrow} 
\newcommand{\adown}{a^\downarrow} 
 
\newcommand{\barr}{\begin{eqnarray}} 
\newcommand{\nd}{\noindent}
\newcommand{\earr}{\end{eqnarray}} 

\begin{document}
\begin{flushright} 
INFNCA-TH0019 \\ 
hep-ph/0010291 \\ 
\end{flushright} 
\vskip 1.5cm 
\begin{center} 
{\bf {\mbox{\boldmath $\Lambda$}} 
polarization in unpolarized hadron reactions\footnote{Talk delivered 
by U.\ D'Alesio at the International Workshop ``Symmetries and Spin''  
Praha-SPIN-2000, July 17-22, 2000, Prague}
}\\ 
\vskip 0.8cm 
{\sf M.\ Anselmino$^1$, D.\ Boer$^2$, U.\ D'Alesio$^3$ and F.\ Murgia$^3$} 
\vskip 0.5cm 
{$^1$ Dipartimento di Fisica Teorica, Universit\`a di Torino and \\ 
      INFN, Sezione di Torino, Via P. Giuria 1, I-10125 Torino, Italy\\ 
\vskip 0.5cm 
$^2$ RIKEN-BNL Research Center\\ 
Brookhaven National Laboratory, Upton, NY 11973, USA\\  
\vskip 0.5cm 
$^3$  Istituto Nazionale di Fisica Nucleare, Sezione di Cagliari\\ 
      and Dipartimento di Fisica, Universit\`a di Cagliari\\ 
      C.P. 170, I-09042 Monserrato (CA), Italy} \\ 
\end{center} 
\vskip 1.5cm 
\noindent 
\noindent 
{\bf Abstract:} \\  
The transverse polarization observed in the inclusive production
of $\Lambda$ hyperons in the high energy collisions of  
{\it unpolarized} hadrons is tackled by considering a new set of  
spin and $\bfk_\perp$ dependent quark fragmentation functions. 
Simple phenomenological expressions for these new 
``{\it polarizing  fragmentation functions}'' 
are obtained by a fit of the data  on $\Lambda$'s  
and $\bar\Lambda$'s produced in $p-N$ processes.

\vskip 18pt
\nd 
{\bf 1. Introduction} 
\vskip 6pt 
$\Lambda$ hyperons produced with   
$x_F \simorder 0.2$ and $p_T \simorder$ 1 GeV/$c$ in the collision of two  
unpolarized  
hadrons, $A B \to \Lup X$, are strongly 
polarized perpendicularly to the production  
plane, as allowed by parity invariance; 
despite 
a huge amount of available experimental information  
on such single spin asymmetries \cite{data}: 
\beq 
P_\Lambda =  
\frac{d\sigma^{A B \to \Lup X} - d\sigma^{A B \to \Ldown X}} 
     {d\sigma^{A B \to \Lup X} + d\sigma^{A B \to \Ldown X}} \>,
\label{pol} 
\eeq 
%
%
no convincing theoretical explanation or understanding of the  
phenomenon exist \cite{theo1}. The perturbative QCD dynamics forbids 
any sizeable single spin asymmetry at the partonic level; 
the polarization of hyperons 
must then originate from nonperturbative features, 
presumably in the hadronization process. 
 
In the last years a phenomenological description of
other single spin asymmetries observed in $\pup p \to \pi X$ reactions  
has been developed
with the introduction of new  
distribution \cite{Ralst-S-79,siv,noi1,dan1} and/or fragmentation  
\cite{col,noi3,Mulders-Tangerman-96}  
functions which are spin and $\bfk_\perp$ dependent; $\bfk_\perp$ denotes 
either the transverse momentum of a quark inside a nucleon or of a hadron 
with respect to the fragmenting quark.

We consider here an effect similar to that suggested by Collins, namely 
a spin and $\bfk_\perp$ dependence in the fragmentation 
of an {\it unpolarized} quark into a {\it polarized} hadron: a function  
describing this mechanism was first introduced in Ref.\  
\cite{Mulders-Tangerman-96} and denoted by $D_{1T}^\perp$. 
More details on this type of definition 
of fragmentation (or decay) functions can be found in Refs.\ 
\cite{col,Mulders-Tangerman-96,elena}.  
 
In the notations of Ref.\ \cite{noi3} a similar function is defined as: 
$
\Delta^N D_{\hup/a}(z, \bfk_{\perp}) \equiv 
\hat D_{\hup/a}(z, \bfk_{\perp}) - \hat D_{\hdown/a}(z, \bfk_{\perp})  
\label{deld1}
= \hat D_{\hup/a}(z, \bfk_{\perp})-\hat D_{\hup/a}(z, - \bfk_{\perp}) \>$, 
and denotes the difference between the density numbers  
$\hat D_{\hup/a}(z, \bfk_{\perp})$ and  
$\hat D_{\hdown/a}(z,$ $\bfk_{\perp})$ 
of spin 1/2 hadrons $h$, with longitudinal momentum fraction $z$, transverse  
momentum $\bfk_{\perp}$ and transverse polarization $\uparrow$ or  
$\downarrow$, inside a jet originated by the fragmentation of an  
unpolarized parton $a$. 

In the sequel we shall refer to $\Delta^N D_{\hup/a}$ and $D_{1T}^\perp$ 
as ``{\it polarizing fragmentation functions}'' \cite{paper}. 

In analogy to Collins' suggestion for the fragmentation of a transversely  
polarized quark \cite{col}, we write: 
\beq  
\hat D_{\hup/q}(z, \bfk_\perp) = \frac 12 \> \hat D_{h/q}(z, k_\perp) +  
\frac 12 \> \Delta^ND_{\hup/q}(z, k_\perp) \>  
\frac{\hat{\bfP}_h \cdot (\bfp_q \times \bfk_\perp)} 
{|\bfp_q \times \bfk_\perp|} \label{lamfn} 
\eeq 
for an unpolarized quark with momentum $\bfp_q$ which fragments into  
a spin 1/2 hadron $h$ with momentum $\bfp_h = z \bfp_q + \bfk_\perp$ 
and polarization vector along the $\uparrow \> = \hat{\bfP}_h$ direction; 
$\hat D_{h/q}(z, k_\perp) 
$ is the $k_\perp$ 
dependent unpolarized fragmentation function, with 
$k_\perp = |\bfk_{\perp}|$. 
From Eq. (\ref{lamfn}) it is clear that the function  
$\Delta^N D_{\hup/a}(z, \bfk_{\perp})$ vanishes in case the transverse  
momentum $\bfk_\perp$ and the transverse spin $\hat{\bfP}_h$ are parallel. 

By taking into account intrinsic $\bfk_\perp$ in the hadronization  
process, and assuming that a QCD factorization theorem holds also when  
$\bfk_\perp$'s are included \cite{col}, one has:
\barr 
\hspace*{-6pt}
\frac{E_\Lambda \, d^3\sigma^{AB \to \Lambda X}}{d^3 \bfp_\Lambda} \>  
P_\Lambda &=& 
\sum_{a,b,c,d} \int \frac{dx_a \, dx_b \, dz}{\pi z^2} \> d^2\bfk_\perp \>  
f_{a/A}(x_a) \> f_{b/B}(x_b) \nonumber \\ 
&\times& \hat s \> \delta(\hat s + \hat t + \hat u) \>  
\frac{d\hat\sigma^{ab \to cd}}{d\hat t}(x_a, x_b, \bfk_\perp) \>  
\Delta^ND_{\Lup/c}(z, \bfk_\perp) \label{phgen} \>.
\earr 

Eq.\ (\ref{phgen}) holds for any spin 1/2 baryon; we shall use it also  
for $\bar\Lambda$'s, with 
$D_{\bar\Lambda/\bar q} = D_{\Lambda/q}$ 
and $\Delta^ND_{\bar\Lambda^\uparrow/\bar q} = \Delta^ND_{\Lup/q}$. 
 
Notice that, in principle, there might be another contribution to  
the polarization of a final hadron produced at large $p_T$ in the high  
energy collision of two unpolarized hadrons; in analogy to   
Sivers' effect \cite{siv, noi1} one might introduce a new spin and 
$\bfk_\perp$ dependent distribution function ($h_1^\perp$ in \cite{dan1}):  
$ 
\Delta^N f_{\aup/A}(x_a, \bfk_{\perp a}) \equiv
\hat f_{\aup/A}(x_a, \bfk_{\perp a}) - \hat f_{\adown/A}(x_a, \bfk_{\perp a})  
\label{delf1}
= \hat f_{\aup/A}(x_a, \bfk_{\perp a}) 
- \hat f_{\aup/A}(x_a, - \bfk_{\perp a}) \>. 
$

We shall not consider this contribution here; not only because of the 
theoretical problems\footnote{The appearence of this function requires initial
state interactions.}
concerning $\Delta^N f_{\aup/A}(x_a, \bfk_{\perp a})$,  
but also because the experimental evidence 
of
the hyperon polarization 
suggests that the mechanism responsible for   
the polarization is in the hadronization process. 
A clean test of this should come from 
a measurement of $P_\Lambda$ in unpolarized DIS  
processes, $\ell p \to \Lup X$ \cite{prep}.    

The main difference between the function $\Delta^N D_{h/a^\uparrow}$ as 
originally proposed by Collins, and the function under present investigation  
$\Delta^N D_{\hup/a}$, is that the former is a so-called chiral-odd function, 
whereas the latter function is chiral-even. Since the pQCD interactions 
conserve chirality, chiral-odd functions must always be accompanied by a  
mass term or appear in pairs.  
Both options restrict the accessibility of such functions. 
On the other hand, the chiral-even fragmentation 
function can simply occur accompanied by the unpolarized (chiral-even) 
distribution functions
allowing for a much cleaner extraction of the fragmentation 
function itself. 

We only consider the quark fragmenting into a $\Lambda$ and use effective 
-- totally inclusive --  unpolarized and  
polarizing $\Lambda$ fragmentation functions to take into account 
secondary $\Lambda$'s  from the decay of other hyperons, like the $\Sigma^0$.  
This  is justified on the basis that
the main $\Sigma^0 \to \Lambda \gamma$ background 
does not produce a significant  
depolarizing effect for the transverse $\Lambda$ polarization.

\vskip 18pt
\nd
{\bf 2. Numerical fits and results} 
\vskip 6pt 

Eq.\ (\ref{phgen}) can be schematically expressed as  
\barr 
&& d\sigma^{pN \to \Lambda X} \> P_\Lambda =  
 d\sigma^{pN \to \Lup X} - d\sigma^{pN \to \Ldown X}   
= \sum_{a,b,c,d} f_{a/p}(x_a) \otimes  f_{b/N}(x_b) \nonumber \\ 
&\otimes& \!\!\!\! [d\hat\sigma^{ab \to cd}(x_a, x_b, \bfk_\perp)  
- d\hat\sigma^{ab \to cd}(x_a, x_b, -\bfk_\perp)] 
\otimes \Delta^ND_{\Lup/c}(z, \bfk_\perp)  
\label{phsch} 
\earr 
which shows clearly that $P_\Lambda$ is a higher twist  
effect, despite the fact that the polarizing fragmentation function 
$\Delta^ND_{\hup/a}$ is a leading twist function: this is due to the  
difference in the square brackets, $[d\hat\sigma(+\bfk_\perp) -  
d\hat\sigma(-\bfk_\perp)] \sim k_\perp/p_T$.
More details can be found in \cite{paper}.

We now use Eq.\ (\ref{phsch}) in order to see whether or not it can reproduce 
the data and in order to obtain information on the new polarizing   
fragmentation functions. To do so we introduce a simple parameterization  
for these functions and fix the parameters by fitting the existing 
data on $P_\Lambda$ and $P_{\bar\Lambda}$ \cite{pl1}-\cite{pl4}. 
 
We assume that $\Delta^ND_{\Lup/c}(z, \bfk_\perp)$ is strongly peaked 
around an average value $\bfk^0_\perp$ lying in the production plane, 
so that we can expect: 
\beq 
\int_{(+k_\perp)} \hskip-18pt d^2\bfk_\perp \>  
\Delta^ND_{\Lup/c}(z, \bfk_\perp) \> F(\bfk_\perp) 
\simeq \Delta_0^ND_{\Lup/c}(z, k^0_\perp) \> F(\bfk^0_\perp) \>. 
\label{delta} 
\eeq 
The average $k^0_\perp$ value depends on $z$ and we parameterize this  
dependence in a most natural way: 
$ 
k^0_\perp(z)/M 
= K \> z^a (1-z)^b \>, 
$
where $M$ is a momentum scale ($M = 1$ GeV/$c$).
 
We parameterize $\Delta_0^ND_{\Lup/c}(z, k^0_\perp)$ in a similar simple  
form but taking into account the positivity condition 
$|\Delta^ND_{\hup/q}(z, k_\perp)| \leq  
\hat D_{h/q}(z, k_\perp)$. 
However, for reasons related to kinematical effects relevant  
at the boundaries of the phase space (see \cite{paper}) 
we prefer to impose the 
more stringent bound $|\Delta_0^ND_{\Lup/c}(z, k^0_\perp)| \leq 
D_{\Lambda/c}(z)/2$,  by taking: 
\beq
\Delta_0^ND_{\Lup/q}(z, k^0_\perp) = 
N_q \, z^{c_q}(1-z)^{d_q} \> \frac {D_{\Lambda/q}(z)}2 \>, 
\label{pard} 
\eeq
where we require $c_q > 0$, $d_q > 0$, and $|N_q|\leq 1$.  

We consider non vanishing contributions in Eq.\ (\ref{pard})
only for $\Lambda$ valence quarks, $u$, $d$ and $s$.
We use the set of unpolarized fragmentation functions of Ref.\ \cite{ind},  
which allows a separate determination of $D_{\Lambda/q}$ and   
$D_{\bar\Lambda/q}$;
in this set the non strange fragmentation  
functions $D_{\Lambda/u} = D_{\Lambda/d}$ are suppressed by an 
$SU(3)$ symmetry breaking factor $\lambda=0.07$ as compared to  
$D_{\Lambda/s}$. In our parameterization of   
$\Delta_0^ND_{\Lup/q}(z, k^0_\perp)$, Eq.~(\ref{pard}), 
we keep the same parameters  
$c_q$ and $d_q$ for all quark flavours, but 
different values of $N_u = N_d$ and $N_s$.

Our best fit results ($\chi^2$/d.o.f. = 1.57) are shown in Figs.~1-3.

\begin{figure}[t]
\begin{center}
  \epsfig{file=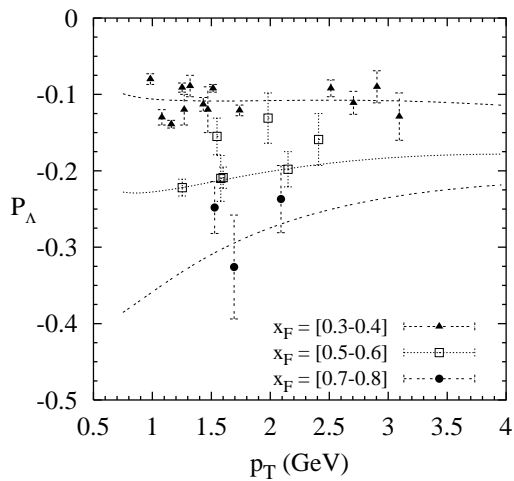,angle=-90,width=7.5cm} 
\end{center}
\vspace{-2mm}
\caption{
   Our best fit to $P_\Lambda$ data from $p$--$Be$ reactions 
   \cite{pl1}-\cite{pl4}  as a function of $p_T$.
   For each $x_F$-bin, the corresponding theoretical curve is evaluated 
   at the mean $x_F$ value in the bin.}
\end{figure}

In Fig.~1 we present our best fits to $P_\Lambda$ as a function of  
$p_T$ for different $x_F$ values, 
as indicated in the figure\footnote{Analogous results have been found 
for the other $x_F$-bins not shown here \cite{paper}.}: the famous 
approximately flat $p_T$ dependence, for $p_T$ greater than 1 GeV/$c$, is 
well reproduced. Such a behaviour, as expected, does not continue 
indefinitely with $p_T$ and we have explicitly checked that at larger values 
of $p_T$ the values of $P_\Lambda$ drop to zero. 
It may be interesting to note that this fall-off has not yet been observed 
experimentally, but is expected to be first seen in the near-future  
BNL-RHIC data.  
Also the increase of $|P_\Lambda|$ with $x_F$ at fixed $p_T$ values 
can be well described, as shown in Fig.~2.

\begin{figure}[t] 
 \begin{center} 
  \epsfig{file=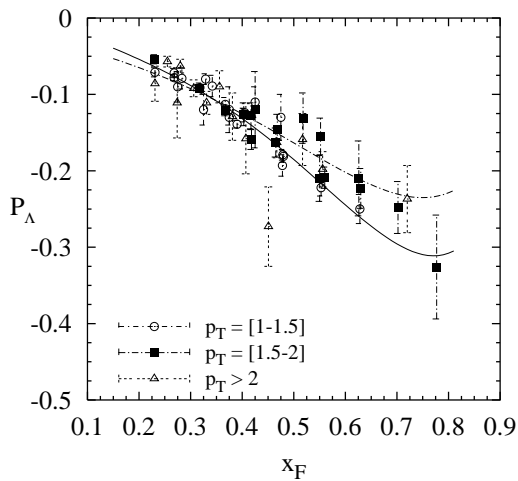,angle=-90,width=7.5cm} 
 \end{center} 
 \vspace{-2mm} 
\caption{
   $P_\Lambda$ data for $p$--$Be$ reactions \cite{pl1}-\cite{pl4}, 
   as a function of $x_F$. 
   The two theoretical curves correspond to $p_T=1.5$ GeV$/c$ (solid) 
   and $p_T=3$ GeV$/c$ 
   (dot-dashed). } 
\end{figure}

\begin{figure}[b!h] 
 \begin{center} 
  \epsfig{file=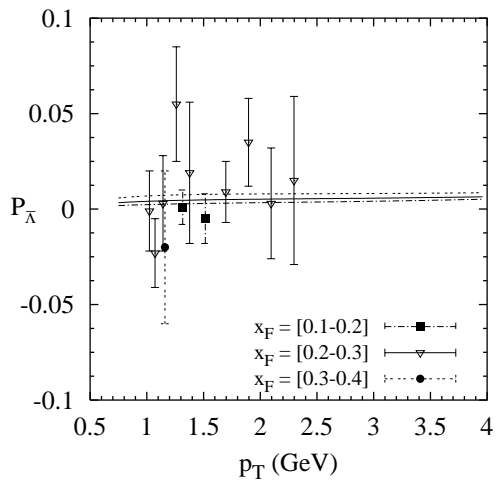,angle=-90,width=7.5cm} 
 \end{center} 
 \vspace{-2mm} 
 \caption{ 
   Our best fit to $P_{\bar\Lambda}$ data from $p$--$Be$ reactions 
   \cite{pl1,pl3}, as a function of $p_T$. 
}
\end{figure} 

Experimental data \cite{pl1}-\cite{pl4} 
are collected at two different c.m. energies, 
$\sqrt{s}\simeq82$ GeV and $\sqrt{s}\simeq116$ GeV. 
Our calculations are performed at  
$\sqrt s = 80$ GeV; we have explicitly checked that by varying the  
energy between 80 and 120 GeV, our results for $P_\Lambda$ vary, in the  
kinematical range considered here, at most by 10\%, in agreement with the  
observed energy independence of the data. 

In Fig.~3  
we show our best fit results for $P_{\bar\Lambda}$ as a function  
of $p_T$ for different $x_F$ values: 
in this case all data \cite{pl1,pl3} are compatible with zero.

The fitted average $k^0_\perp$ value of a $\Lambda$ inside a jet 
turns out to be very reasonable: $ K = 0.69$, $a = 0.36$ and $b = 0.53$. 
Also, mostly $u$ and $d$ quarks 
contribute to $P_\Lambda$, resulting in a negative value of $N_u$; instead, 
$u$, $d$ and $s$ quarks all contribute significantly 
to $P_{\bar\Lambda}$ 
and opposite signs for $N_u$ and $N_s$ are found, inducing cancellations.

We have also considered  a second -- $SU(3)$ symmetric --
set of fragmentation functions  $D_{\Lambda/q}$ \cite{defl}. 
One reaches similar  
conclusions about the polarizing fragmentation functions 
$\Delta^ND_{\Lup/q}$: $N_{u,d} \not= N_s$ and not only is there a difference 
in magnitude, but once more one finds negative values for 
$\Delta^N_0D_{\Lup/u,d}$ and positive ones for $\Delta^N_0D_{\Lup/s}$. 
This seems to be a well established general trend \cite{paper}. 

\vskip 18pt 
\nd
{\bf 3. Conclusions} 
\vskip 6pt 
 
We have considered here the well known and longstanding problem of the  
polarization of $\Lambda$ hyperons, produced at large $p_T$ in the  
collision of two unpolarized hadrons in  a generalized  
factorization scheme -- with the inclusion of intrinsic transverse  
motion -- with pQCD dynamics. 
The new, spin and $\bfk_\perp$ dependent, 
polarizing fragmentation functions $\Delta^ND_{\Lup/q}$ have been  
determined by a fit of 
data on $p \, Be \to \Lup X$,  
$p \, Be \to \bar\Lambda^\uparrow X$ and $p\,p \to \Lup X$.

The data can be described with remarkable accuracy in all their features:
the large negative values of the $\Lambda$ polarization, the increase of
its magnitude with $x_F$, the puzzling flat $p_T \simorder 1$ GeV/$c$ 
dependence and the $\sqrt s$ independence; 
also the 
tiny or zero values of $\bar\Lambda$ polarization are well reproduced. 

Our parameterization of $\Delta^ND_{\Lup/q}$ should allow us 
to give predictions for $\Lambda$ polarization in other processes; a study of  
$\ell p \to \Lup \, X$, $\ell p \to \ell' \Lup \, X$ 
 and $e^+e^- \to \Lup \, X$ is in progress \cite{prep}.  

\vskip 18pt  
\nd   
{\bf Acknowledgements} 
\vskip 6pt
Two of us (U. D. and F. M.) are partially supported by COFINANZIAMENTO 
MURST-PRIN.

\end {document}